\documentclass[graybox]{svmult}

\usepackage{type1cm}        
\usepackage{makeidx}         
\usepackage{graphicx}        
\usepackage{multicol}        
\usepackage[bottom]{footmisc}

\usepackage{newtxtext}       %
\usepackage{newtxmath}       

\usepackage{natbib}
\newcommand{\be}{\begin{equation}}
\newcommand{\ee}{\end{equation}}
\newcommand{\bea}{\begin{eqnarray}}
\newcommand{\eea}{\end{eqnarray}}
\newcommand{\bdm}{\begin{displaymath}}
\newcommand{\edm}{\end{displaymath}}
\newcommand{\X}{{\mathbf x}}
\newcommand{\Y}{{\mathbf y}}

\makeindex             

\begin{document}

\title*{Space, Time, Matter in Quantum Gravity}

\author{Claus Kiefer}

\institute{Claus Kiefer \at Institute for Theoretical Physics,
  University of Cologne, Z\"ulpicher Stra\ss e 77a, 50937 K\"oln, Germany
  \email{kiefer@thp.uni-koeln.de}} 

\maketitle

\abstract{The concepts of space, time, and matter are of central
  importance in any theory of the gravitational field. Here I
  discuss the role that these concepts might play in quantum theories
  of gravity. To be concrete, I will focus on the
  most conservative approach, which is quantum geometrodynamics.
  It turns out that spacetime is absent at the most fundamental
  level and emerges only in an appropriate limit. It is expected that
  the dynamics of matter can only be understood from a fundamental
  quantum theory of all interactions.}

\section{From classical to quantum gravity}

In his famous habilitation colloquium on June 10, 1854, Bernhard
Riemann concluded

\begin{quote}

The question of the validity of the hypotheses of geometry in the
infinitely small is bound up with the
question of the ground of the metric relations of space.\ldots
Either therefore the reality which underlies space must form a
discrete manifoldness, or we must seek the ground of its metric
relations outside it, in binding forces which act upon it. \ldots This
leads us into the domain of another science, of physic, into which the
object of this work does not allow us to go to-day. (\cite{Riemann1854};
translated by William Kingdon Clifford~1873)\footnote{The German
  original reads (\cite{JostD}, p.~43):
  ``Die Frage \"uber die G\"ultigkeit der Voraussetzungen der Geometrie im
Unendlichkleinen 
h\"angt zusammen mit der Frage nach dem innern Grunde der
Massverh\"altnisse des Raumes. \ldots
Es muss also entweder das dem Raume zu Grunde liegende Wirkliche
eine discrete Mannigfaltigkeit bilden, oder der Grund der
Massverh\"altnisse ausserhalb, in darauf wirkenden bindenden
Kr\"aften, gesucht werden. \ldots Es f\"uhrt dies hin\"uber in das
Gebiet einer andern Wissenschaft, in das Gebiet der Physik, welches
wohl die Natur der heutigen Veranlassung nicht zu betreten erlaubt.''
The English translation can be found in \cite{JostE}. For the role of
Clifford in the development of these ideas, see e.g.
\cite{Giulini18}.}  

\end{quote}

Riemann's pioneering ideas are important for at least two
reasons. First, although Riemann did not take into account the
time dimension, his ideas led to the mathematical formalism that enabled
Albert Einstein to formulate his theory of general relativity (GR) in
1915. In GR, gravity is understood as the manifestation of a dynamical
geometry of space and time, which are unified into a four-dimensional spacetime.

Second, as is clear from the sentences quoted above, matter and
geometry are no longer imagined as independent from each other; the
metric now depends on the ``binding forces which act upon it''. The
metrical field is no longer given rigidly once and for all, but stands
in causal dependence on matter. This idea is at the core of
in GR. 

In his commentary on Riemann's text from 1919, Hermann Weyl emphasized
the unification of geometry and field theory in physics,

\begin{quote}

For geometry, here the same step happened that Faraday and Maxwell
performed within physics, in particular electricity theory, which was
done by the transition from an action-at-a-distance to a local-action
theory: carrying out the principle to understand the world from its
behaviour in the infinitely small. See
\cite{JostD}, p.~45.\footnote{``F\"ur die Geometrie
  geschah hier der gleiche Schritt, den \textsc{Faraday} und
  \textsc{Maxwell} innerhalb der Physik, speziell der
  Elektrizit\"atslehre, vollzogen durch den \"Ubergang von der
  Fernwirkungs- zur Nahewirkungstheorie: das Prinzip, die Welt aus
  ihrem Verhalten im Unendlichkleinen zu verstehen, gelangt zur Durchf\"uhrung.''}

 \end{quote} 

 Riemann's approach turned out to be much more powerful than
 alternative ideas on the foundation of geometry, for example those of
 Hermann von Helmholtz, see 
 e.g. \cite{JostE}, p.~119. Helmholtz starts from
 experience\footnote{The title of Helmholtz's article, ``Ueber die
   Thatsachen, die der Geometrie zu Grunde liegen'' (``On
   the Facts which Lie at the Bases of Geometry''), makes a dig at
   the title of Riemann's work.} and postulates the possibility of free motion of
 bodies. As he can prove mathematically, for this free motion a space with
 constant curvature is required. From the later perspective of GR,
 this turns out to be too narrow. Riemann's idea, that bodies
 can carry geometry with them, {\em is} realized in GR, which allows
 spaces, in fact spacetimes, to have arbitrary curvature, as
 determined by the Einstein field equations. These equations read

 \be
\label{EFE}
 R_{\mu\nu}-\frac12g_{\mu\nu}R+\Lambda g_{\mu\nu}=\kappa T_{\mu\nu}.
 \ee

 Here, $g_{\mu\nu}$ denotes the spacetime metric, $R_{\mu\nu}$ the
 Ricci tensor, and $R$ the Ricci scalar. Non-gravitational degrees of
 freedom (for simplicity called `matter') are described by a symmetric
 energy--momentum tensor $T_{\mu\nu}$; it obeys the covariant
 conservation law
 \be
 \label{conservation}
T_{\mu\nu;}^{\ \ \ \nu}=0.
\ee
It is important to emphasize that this is not a standard conservation law
(with a partial instead of a covariant derivative) from which a
conserved current and charge can be 
derived.\footnote{This is only possible in the presence of a symmetry,
  as expressed by a Killing
  vector.} If the energy--momentum tensor obeys the dominant energy
condition (energy densities dominate over pressures), causality is
implemented in the sense that no influence from outside the lightcone
can enter its inside. 

 There are two free parameters in the gravitational sector: $\kappa$
 and $\Lambda$. From the Newtonian limit, one can identify
 \be
 \label{kappa}
 \kappa=8\pi G/c^4,
 \ee
 with $G$ the gravitational (Newton) constant and $c$ the
 speed of light. In 1917, Einstein had recognized that another
 free parameter is allowed -- the cosmological constant $\Lambda$
 which has the physical dimension of an inverse length squared.
 From observations we find the value $\Lambda\approx 1.2\times
 10^{-52}{\rm m}^{-2}\approx 0.12\ ({\rm Gpc})^{-2}$.\footnote{Recent
   doubts on this $\Lambda$-observation are expressed e.g. in
   \cite{VMS20}.}
The relation of this value to naive estimates from quantum field
theory is an open question. 

 The Einstein field equations \eqref{EFE} describe a non-linear
 interaction between geometry and matter. In this sense, $T_{\mu\nu}$
 must not be interpreted as the source from which the metric is
 determined. For the description of matter, the metric is also needed,
 since it enters the field equations for matter as well as the equation of
 motion for test bodies\footnote{Test bodies in GR cannot be mass
   points. The mass-to-radius ratio of objects has an upper bound
 of $c^2/2G$; the concept of a mass point is replaced by a black
 hole.} given by
 \be
 \label{geodesic}
\ddot{x}^{\mu}+\Gamma^{\mu}_{\alpha\beta}\dot{x}^{\alpha}\dot{x}^{\beta}=0.
  \ee
  Here, $\Gamma^{\mu}_{\alpha\beta}$ are the components of the
  Levi--Civita connection, which is determined by the metric, and the
  dots denote derivatives with respect to proper time (for timelike
  geodesics) or with respect to an affine parameter (null geodesics).  Equation
  \eqref{geodesic} is the geodesic equation which reflects the
  universal coupling of gravity to matter. In contrast to its
  Newtonian analogue, it corresponds to 
  free motion in the geometry described by $g_{\mu\nu}$ (`equivalence
  principle').
  So for the description of matter, the pair
  $(T_{\mu\nu},g_{\alpha\beta})$ is needed, and one needs a rather
  involved initial value formulation to determine the spacetime metric
  (see next section).

  A major feature of GR, and one that is particularly relevant for its
  quantization, is {\em background independence}. This must be
  carefuly distinguished from mere general covariance, which means
  form invariance of equations under an arbitrary change of
  coordinates.  In contrast, background independence means that there are no
  absolute (non-dynamical) fields in the theory -- this applies to GR,
  where the metric is a dynamical quantity that acts on matter and is
  acted upon by it. As J\"urgen Ehlers has remarked (\cite{Ehlers}, p.~91):
``Conceptually, the background 
independence must be seen as the principal achievement of general
relativity theory; it is, however, at the same time the main obstacle
to overcome if general relativity theory and quantum theory are to be
united.'' In GR, the law of motion \eqref{geodesic} cannot be
formulated independently from the field equations \eqref{EFE} -- 
in fact, it follows from them by employing \eqref{conservation}. This would not
be possible in a theory with an absolute background, that is, with an
absolute non-dynamical spacetime.

In 1918, Weyl generalized the notion of the Levi-Civita connection
that occurs in \eqref{geodesic} to a symmetric linear
connection, \cite{Weyl18}, see also the extended discussion in {\em Raum, Zeit,
  Materie}, \cite{RZM}.
For this concept, a metrical structure on the manifold is not needed, only the
notion of a parallel transport for vectors and tensors, which provides
the means to connect different points on the 
manifold. In contrast to the Levi-Civita connection, his more general
connection need not be derivable from a metric. Weyl distinguishes, in
fact, three levels of geometry: the first level is the topological manifold (which
he calls {\em situs manifold or empty world}),\footnote{{\em Analysis
    situs} is an older name for topology.} the second level the
affinely connected manifold, and the third level the metric continuum
(which he also calls ``ether''); see also \cite{Schrodinger} for a
lucid presentation. 

The notion of a symmetric linear connection allowed Weyl to construct a
generalization of Einstein's theory. In his theory, the magnitude of
vectors is not fixed, but the connection allows the comparison of
magnitudes in different points. This introduces a new freedom into the
theory -- the freedom to perform gauge transformations. The metric is
here determined only up to a (spacetime-dependent) factor. The
exponent of this factor can be connected with a function that behaves
as the electromagnetic vector potential (here interpreted as a
one-form). Weyl thought that he has 
constructed in this way a unified theory of gravity and
electromagnetism; for details, see \cite{RZM},
p.~121~ff.\footnote{Here, the words `gauge' ({\em Eichung}), `to
  gauge' ({\em eichen}), and `gauge invariance' ({\em Eich-Invarianz})
  enter. Their original meaning arises from providing
  standards for physical quantities (including distances), which is
  different from their later abstract use in the
  description of intrinsic symmetries in gauge theories.} 
In the above hierarchy, Weyl's
theory can be located between the second and third level: in it, spacetime has a
conformal structure, which provides a more general framework than the structure of
Riemannian geometry.

Weyl was convinced that fundamental geometric relations
should only refer to infinitesimally neighbouring points ({\em
  Nahgeometrie} instead of {\em Ferngeometrie}). This principle plays
a key role in both the 1918 and the 1929 versions of gauge theories.
In \cite{Weyl-Ph}, p.~115, he writes (emphasis by Weyl): ``{\em Only in the
infinitely small can we expect to encounter everywhere the same elementary
laws}, thus the world must be understood from its behaviour in the
infinitely small.\footnote{The German original reads: ``{\em Nur im
    Unendlichkleinen d\"urfen wir erwarten auf die elementaren,
    \"uberall gleichen Gesetze zu sto\ss en}, darum mu\ss\ die Welt
  aus ihrem Verhalten im Unendlichkleinen verstanden werden.}

In spite of its formal elegance, Weyl's theory is empirically wrong,
as was soon realized by Einstein. The reason is that a non-integrable
connection leads to path-dependent frequencies for atomic spectra, in
contrast to observations.
But his theory can nevertheless be seen as the
origin of our modern gauge theories. A decade later, in 1929, Weyl came up
with a gauge theory of electromagnetism and the Dirac field. Instead
of the real conformal factor multiplying the metric, there occurs now
a phase factor for which the exponent is a one-dimensional integral
over the vector potential. A non-integrable connection is manifested
there, for example, in the Aharonov--Bohm effect.
In its non-Abelian generalization, gauge
invariance is a key ingredient to the Standard Model of particle
physics, see e.g. \cite{Dosch} for a review. The Standard Model
(extended by massive neutrinos) is
experimentally extremely well tested, and no obvious deviation from it is seen
so far in experiments at the Large Hadron Collider (LHC) and elsewhere.

But what about gravity and spacetime? In its standard formulation,
GR is not a gauge theory. The reason
is that the connection $\Gamma^{\mu}_{\alpha\beta}$ is not independent
there, but is
derived from a metric. On thus has the chain 
\bdm
g_{\mu\nu}\longrightarrow \Gamma^{\mu}_{\alpha\beta} \longrightarrow
R^{\mu}_{\alpha\beta\gamma},
\edm
where $R^{\mu}_{\alpha\beta\gamma}$ denotes the Riemann curvature tensor. For
gauge theories, the first step in this chain is lacking. Gauge
theories of gravity do, however, exist, and they are needed for the
consistent implementation of fermions, see \cite{reader}.\footnote{See
  also the contributions by Hehl and Obukhov and by Scholz to this
  volume.} Weyl's original theory is a special case of this general
class, but it is important to emphasize that the coupling of Weyl's
vector potential is not to the electrodynamic current -- as its
creator believed -- but to the dilaton current (because the
one-parameter dilation group is gauged), see \cite{HCM88}. 

One of the striking properties of GR is that it exhibits its own
incompleteness. This is expressed in the singularity theorems which
state that, under general conditions, singularities in spacetime are
unavoidable, see \cite{HP96}. Singularities are here understood in the
sense of geodesic incompleteness -- timelike or null geodesics as found from
\eqref{geodesic} terminate at finite proper time or finite affine
parameter value. In most physically 
relevant cases, the occurrence of singularities is connected with
regions of infinite curvature or energy density; notable examples are
the singularities characterizing the beginning of the Universe (``big
bang'') and the interior of black holes. One of the hopes connected
with the construction of a quantum theory of gravity is that such a
theory will avoid singularities. This hope may be extended to
a different type of singularities in our present physical theories --
the infinities that arise in 
almost every local quantum field theory. One has learnt to cope with the
latter singularities by employing sophisticated methods of
regularization and renormalization. Nevertheless, one would expect
that a truly fundamental
theory will be finite from the onset. The reason is that the occurrence of
singularities is connected with an unsufficient understanding of the
microstructure of spacetime. True infinities should not occur in any
sensible description of Nature, cf. \cite{EMN18}. 

One possible solution to the singularity problem is to avoid a
continuum for the spacetime structure and to assume instead that
spacetime is built up from discrete entities. There are indications for
such a discrete structure in some approaches to quantum
gravity, but the last work has not yet been spoken. Interestingly,
Riemann himself envisaged the possibility of a continuous as well as a
discrete manifold; the smallest entities he calls {\em quanta}
(\cite{JostE}, p.~32):

\begin{quote}
Definite portions of a manifoldness, distinguished by a mark or by a
boundary, are called Quanta. Their comparison with regard to quantity
is accomplished in the case of discrete magnitudes by counting, in the
case of continuous magnitudes by measuring. \\ (Translated by William
Kingdon Clifford~1873)\footnote{The German original reads
  (\cite{JostD}, p.~ 31):
  ``Bestimmte, durch ein Merkmal oder eine Grenze unterschiedene
  Theile einer Mannigfaltigkeit heissen Quanta. Ihre Vergleichung der
  Quantit\"at nach geschieht bei den discreten Gr\"ossen durch
  Z\"ahlung, bei den stetigen durch Messung.''}
 \end{quote} 
Weyl, in his commentary to Riemann's text, speculates that the final
answer to the problem of space may be found in its discrete
nature.\footnote{``Sehen wir von der ersten M\"oglichkeit ab, es
  k\"onnte `das dem Raum zugrunde liegende Wirkliche eine diskrete
  Mannigfaltigkeit bilden' ({\em obschon in ihr vielleicht einmal die
    endg\"ultige Antwort auf das Raumproblem enthalten sein wird}, my
  emphasis) \ldots}

What happens to this when the quantum of action $\hbar$ comes into
play? One of the early pioneers of attempts to quantizing gravity,
Matvei Bronstein, through the application of thoughts experiments,
arrived at the necessity 
of introducing minimal distances in spacetime, thus abandoning the
idea of a metric continuum. He writes\footnote{The quotation is from
  \cite{OUP}, p.~20.}
\begin{quote}

The elimination of the logical inconsistencies connected with this
[his thought experiments] requires a radical reconstruction of the
theory, and in particular, 
the rejection of a Riemannian geometry dealing, as we see here,
with values unobservable in principle, and perhaps also the
rejection of our ordinary concepts of space and time, modifying
them by some much deeper and nonevident concepts.
{\em Wer's nicht glaubt, bezahlt einen Taler}.

 \end{quote} 

 In Bronstein's analysis, quantities appear that can be found by
 combining $G$, $c$, and $\hbar$ into units of length, time, and mass
 (or energy). They were first presented by Max Planck in 1899 (one
 year before the `official' introduction of the quantum of action into
 physics!) and are called {\em Planck units} in his honour. They
 read\footnote{See e.g. \cite{OUP}, p.~5.}
 \begin{eqnarray}
l_{\rm P} &=& \sqrt{\frac{\hbar G}{c^3}} \approx 1.62\times 10^{-35}\ 
{\rm m}
\\
t_{\rm P} &=& \frac{l_{\rm P}}{c}=\sqrt{\frac{\hbar G}{c^5}}
\approx 5.40\times 10^{-44}\ {\rm s}
\\
m_{\rm P} &=& \frac{\hbar}{l_{\rm P}c}=\sqrt{\frac{\hbar c}{G}}
\approx 2.17\times 10^{-8}\ {\rm kg}\approx 1.22 \times 10^{19}\ {\rm GeV}/c^2.
 \end{eqnarray}
 At the end of his 1899 paper, Planck wrote the following prophetic
 sentences, see \cite{OUP}, p.~6:

\begin{quote}
These quantities retain their natural meaning as long as
the laws of gravitation, of light propagation in vacuum, and the two laws of 
the theory of heat remain valid; they must therefore, if measured 
in various ways by all kinds of intelligent beings, always turn out to be the 
same.\footnote{The German original reads:
  ``Diese Gr\"ossen 
behalten ihre nat\"urliche Bedeutung
so lange bei, als die Gesetze der Gra\-vi\-ta\-tion, der Lichtfortpflanzung
im Vacuum und die beiden Haupts\"atze der W\"armetheorie in
G\"ultigkeit bleiben, sie m\"ussen also, von den verschiedensten
Intelligenzen nach den verschiedensten Methoden gemessen,
sich immer wieder als die n\"amlichen ergeben.''}
\end{quote}

One can form a dimensionless number out of these Planck units by
bringing the cosmological constant $\Lambda$ into play. Inserting the
present observational value for $\Lambda$ (see above), this gives
\be
l_{\rm P}^2\Lambda\equiv \frac{G\hbar\Lambda}{c^3}\approx 3.3\times10^{-122}.
\ee
The smallness of this number is one of the biggest open puzzles in
fundamental physics. Only a fundamental unified theory of all interactions is
expected to provide a satisfactory explanation.

What are the general arguments that speak in favour of a quantum
theory of gravity?\footnote{See e.g. \cite{OUP} for a
  comprehensive discussion.} First, as mentioned above, there is the
  singularity problem of 
  classical general relativity, which points to the incompleteness of
  Einstein's theory. Second, the search for a unified theory of all
  interactions should include quantum gravity: gravity interacts
  universally to all fields of Nature, and all non-gravitational
  fields are successfully described by quantum (field) theory so far,
  so a quantum description should apply to gravity, too. Third, a very general argument
  was put forward by Richard Feynman in 1957, see \cite{OUP}, p.~18:
  if we generate a superposition of two masses at different
  locations, their gravitational fields should also be superposed,
  unless the superposition principle of quantum theory breaks down. A
  quantum theory of gravity is needed to describe such
  superpositions. It is clear that such a state can no longer correspond
  to a classical spacetime. There are at present interesting 
  suggestions for the possibility to observing the
  gravitational field generated by a 
  quantum superposition in laboratory experiments, see \cite{CBPU19} and
   references therein.  
 
Several approaches to quantum gravity exist, but there is so far no consensus
in the community, see \cite{OUP}. The ideal case would be to construct
a finite quantum theory of all interactions from which present
physical theories can be derived as approximations
(or ``effective field theories'') in appropriate limits. The only
reasonable candidate is string theory. In this theory, the dimension
of spacetime assumes the number ten or eleven.
Unfortunately, it is so far not
clear how to recover the Standard Model from string theory and how to
test it by experiments. Connected with this is the difficulty to
proceed in a more or less unique way from the ten or eleven spacetime
dimensions to the four dimensions of the observed world. 

The main alternatives to finding a unified theory are the more modest
attempts to construct first a quantum theory of the gravitational field
and to relegate unification to a later step. The usual starting point
is GR, but quantization methods may be applied to any other
gravitational theory. Standard methods are path integral
quantization and canonical quantization. We shall focus below on the
canonical quantization of GR using metric variables, because
conceptual issues dealing with 
space and time are most transparent in this approach, see \cite{CK09}.


\section{The configuration space of general relativity}

Besides ordinary three-dimensional space (or four-dimensional
spacetime), the concept of {\em configuration space} plays an eminent
role in physics. In mechanics, this is the $N$-dimensional space generated by all
configurations, described by coordinates $\{ q^a\}, \ a=1,\ldots N$,
that the system can assume. In field theory, it is
infinite-dimensional of possible field configurations. In quantum
theory, it will enter the argument of the wave function (functional)
and lead to the central property of entanglement. 

What is the configuration space in general relativity?
As John Wheeler writes (\cite{Battelle}, p.~245): ``A decade and more
of work by Dirac, Bergmann, Schild, Pirani, Anderson, Higgs, Arnowitt,
Deser, Misner, DeWitt, and others has taught us through many a hard
knock that Einstein's geometrodynamics deals with the dynamics of
geometry: of 3-geometry, not 4-geometry.'' Most of these developments
happened after Weyl's death in 1955. In fact, upon application of the
canonical (or Hamiltonian) formalism, Einstein's theory
can be written as a dynamical system for the {\em three-metric}
$h_{ab}$ and its canonical momentum $\pi^{ab}$ on a spacelike
hypersurface $\Sigma$. The ten Einstein equations can be formulated as
four constraints, that is, restrictions on initial data $h_{ab}$ and
$\pi^{ab}$ on $\Sigma$, and six evolution equations. The four
constraints read (per spacepoint)
\begin{alignat}{2}
& {\mathcal H}_{\perp} &&\,=\,2\kappa\,G_{ab\,cd}\pi^{ab}\pi^{cd}
\,-\,(2\kappa)^{-1}\sqrt{h}({}^{\scriptscriptstyle (3)}\!R-2\Lambda)
+\,\sqrt{h}\rho\approx 0\label{H-perp}\\
&  {\mathcal H}^a &&\,=-2\nabla_b\pi^{ab}\,+\,\sqrt{h}j^a\approx 0,\label{H-a}
\end{alignat}
with the (inverse) DeWitt metric
\be
\label{DeWitt-inverse}
G_{ab\,cd}=\tfrac{1}{2\sqrt{h}}(h_{ac}h_{bd}+h_{ad}h_{bc}-h_{ab}h_{cd})
\ee
and $\kappa$ given by \eqref{kappa}.
Here, ${}^{\scriptscriptstyle (3)}\!R$ denotes the three-dimensional
Ricci scalar and $h$ the determinant of $h_{ab}$; $\rho$ and $j^a$
denote matter density and current, respectively. 
The constraint ${\mathcal H}_{\perp}\approx 0$ is called ``Hamiltonian constraint'',
while ${\mathcal H}^a\approx 0$ are 
called ``momentum (diffeomorphism) constraints''. The symbol $\approx
0$ is Dirac's weak equality and means ``vanishing as a
constraint''. The canonical momentum $\pi^{ab}$ is related to the
extrinsic curvature $K_{cd}$ of $\Sigma$ by
\be
\label{DeWitt}
\pi^{ab}=\frac{G^{ab\,cd}K_{cd}}{2\kappa},
\ee
where $G^{ab\,cd}$ denotes the DeWitt metric itself (the inverse of
the expression in \eqref{DeWitt-inverse}).
This quantity plays the role of a metric in the space of all
Riemannian three-metrics $h_{ab}$, a space called Riem~$\Sigma$.

Is Riem~$\Sigma$ the configuration space of GR? Not yet. The
constraints ${\mathcal H}^a\approx 0$ guarantee the invariance of the
theory under infinitesimal three-dimensional coordinate
transformations. The real configuration space is thus the space of all
three-{\em geometries}, not the space of all three-{\em metrics}. This
is what Wheeler called {\em superspace}, here denoted by ${\mathcal S}(\Sigma)$, see
\cite{Battelle}. It is 
the arena for classical and quantum geometrodynamics.  One can formally write
\bdm
{\mathcal S}(\Sigma):={\rm Riem}\ \Sigma/{\rm Diff}\ \Sigma,
\edm
where ${\rm Diff}\ \Sigma$ denotes the group of three-dimensional
diffeomorphisms (``coordinate transformations'').
By going to superspace, the momentum constraints are
automatically fulfilled. Whereas ${\rm Riem}\ \Sigma$ has a simple
topological structure, the topological structure
of ${\mathcal S}(\Sigma)$ is very complicated because it
inherits (via ${\rm Diff}\ \Sigma$) some of the topological
information contained in $\Sigma$; see \cite{Giulini09} for
details.

The DeWitt metric has pointwise a Lorentzian signature with one
negative and five positive directions, that is, it has negative, null,
and positive directions. Due to the minus sign, the kinetic term for
the gravitational field is indefinite. It is important to note that
this minus sign is unrelated to the signature of spacetime; starting
with a four-dimensional Euclidean space instead of a four-dimensional
spacetime, the same signature for the DeWitt metric is found. The
presence of this minus sign is related to the attractive nature of
gravity. It is also worth mentioning that the DeWitt metric reveals a
surprising analogy with the elasticity tensor in three-dimensional
elasticity theory and the local and linear constitutive tensor in
four-dimensional electrodynamics, see \cite{HK18}. This analogy could
be of importance for theories of emergent gravity. 

Constraints and evolution equations have an intricate relationship;
see e.g. \cite{GK07}. Let me summarize the main features as well as
pointing out analogies with electrodynamics. First, there
is an important connection with the (covariant) conservation law of
energy--momentum. The constraints are preserved in time if and only if
the energy--momentum tensor of matter has vanishing covariant divergence.
In electrodynamics, the Gauss constraint is preserved in time
if and only if electric charge is conserved.

Second, Einstein's equations represent the unique propagation law
consistent with 
the constraints. To be more concrete, if the constraints are valid on
an ``initial'' hypersurface and if the dynamical evolution equations
(the pure spatial components of the Einstein equations) hold, the
constraints hold on every hypersurface. And if the constraints hold on
every hypersurface, the dynamical evolution equations hold.
Again, there is an analogy with electrodynamics: 
Maxwell's equations are the unique propagation law consistent with
the Gauss constraint.

It must be emphasized that the picture of a spacetime foliated by
a one-parameter family of hypersurfaces
only emerges {\em after} the dynamical equations are
solved. Then, spacetime can be interpreted as a ``trajectory of
spaces''. 
Before this is done, one only has a three-dimensonal manifold
$\Sigma$ with given topology, equipped with the canonical variables
satisfying the constraints \eqref{H-perp} and \eqref{H-a}.

This fact that spacetime is not given from the outset but must be
constructed through an initial value formulation, is an expression of
the background independence discussed in the previous section. In this
sense, the analogy with electrodynamics on a given external spacetime
breaks down. Background independence is related with the classical
version of what is called the {\em problem of time}: if we restrict ourselves to
compact three-manifolds $\Sigma$, the total Hamiltonian of GR is a
combination of the constraints \eqref{H-perp} and
\eqref{H-a}.\footnote{In the asymptotically flat case, additional
  boundary terms are present.} Thus, no external time parameter
exists; all physical time 
parameters are to be constructed from within our system, that is, as 
functional of the canonical variables. A priori,
there is no preferred choice of such an intrinsic time parameter.
It is this absence of an 
external time and the non-preference of an intrinsic one that is known
as the problem of time in (classical) canonical gravity.
Still, after the solution of the dynamical equations, spacetime as a
trajectory of spaces exists. This is different in the quantum theory
where it leads to the far-reaching quantum version of the problem of
time (next section). 

The possibility of constructing spacetime in the way just described,
is also reflected in the closure of the Poisson algebra for the constraints
\eqref{H-perp} and \eqref{H-a}:
\bea
\{{\mathcal H}_{\perp}({\mathbf x}),{\mathcal H}_{\perp}(\Y)\}
&=& -\sigma\delta_{,a}(\X,\Y)\left(h^{ab}(\X){\mathcal H}_b(\X)
+h^{ab}(\Y){\mathcal H}_b(\Y)\right) \label{algebra-1}\\
\{{\mathcal H}_a(\X),{\mathcal H}_{\perp}(\Y)\} &=& {\mathcal
  H}_{\perp}(\X)\delta_{,a}(\X,\Y) \label{algebra-2}\\ 
\{{\mathcal H}_a(\X),{\mathcal H}_b(\Y)\} &=&
 {\mathcal H}_b(\X)\delta_{,a}(\X,\Y)
+{\mathcal H}_a(\Y)\delta_{,b}(\X,\Y)\label{algebra-3}
\eea
It is not a Lie algebra, though, because the Poisson bracket between
two Hamiltonian constraints at different points also contains (the inverse
of) the three-metric, $h^{ab}$. We also remark that the signature of
the spacetime metric enters here in the form of the parameter
$\sigma$: in fact, $\sigma=-1$ corresponds to a four-dimensional spacetime,
while $\sigma=1$ corresponds to a four-dimensional space. It is a
fundamental (and open) question whether the closure of this algebra
also holds in quantum gravity. 

The relation of the transformations generated by the
constraints to the spacetime diffeomorphisms is a subtle one and will
not be discussed here; see e.g. \cite{Sundermeyer,OUP}.


\section{Quantum geometrodynamics}

In the last section, we have reviewed the canonical (Hamiltonian)
formulation of GR. Here, we discuss the quantum version of this, see
e.g. \cite{OUP} for a comprehensive treatment. We follow Dirac's
heuristic approach and transform the classical constraints
\eqref{H-perp} and \eqref{H-a} into conditions on physically allowed
wave functionals. These wave functionals are defined on the space of
all three-metrics (the above space Riem $\Sigma$) and matter fields
on $\Sigma$. The quantum version of \eqref{H-perp} reads
\bea
\label{WDW}
\hat{{\mathcal H}}_{\perp}\Psi&\equiv&
\left(-16\pi G\hbar^2G_{abcd}\frac{\delta^2}{\delta h_{ab}\delta
    h_{cd}}\right.
\nonumber \\
& & \left.-(16\pi G)^{-1}\,\sqrt{h}\bigl(\,{}^{(3)}\!R-2\Lambda\bigr)
+\sqrt{h}\hat{\rho}\right)\Psi=0
\eea
and is called the {\em Wheeler--DeWitt equation}. We note that the
kinetic term in this equation only has formal meaning before the
issues of factor ordering and regularization are successfully
addressed.\footnote{For a recent attempt into this direction, see
  \cite{feng18}.} The quantum implementation of \eqref{H-a} reads
\be
\label{diffeo}
\hat{{\mathcal H}}^a\Psi \equiv -2\nabla_b\frac{\hbar}{\rm i}
\frac{\delta\Psi}{\delta h_{ab}}+\sqrt{h}\hat{j}^a\Psi =0
\ee
and is called the momentum (or diffeomorphism) constraints. These
latter equations have a simple interpretation. 
Under a coordinate transformation 
\bdm
x^a \mapsto \bar{x}^a=x^a+\delta N^a(\X),
\edm
the three-metric transforms as
\bdm
h_{ab}(\X)\mapsto \bar{h}_{ab}({\X})=h_{ab}(\X)-D_a\delta N_b(\X)
-D_b\delta N_a(\X).
\edm
The wave functional then transforms according to
\bdm
\Psi[h_{ab}]\mapsto \Psi[h_{ab}]-2\int\D^3x\ \frac{\delta\Psi}
{\delta h_{ab}(\X)}D_a\delta N_b(\X).
\edm
Assuming the invariance of the wave functional under this
transformation, one is led to
\bdm
D_a\frac{\delta\Psi}{\delta h_{ab}}=0.
\edm
This is exactly \eqref{diffeo} (restricted here to the vacuum case). 

A simple analogy to \eqref{diffeo} is Gauss's law in quantum
electrodynamics (or its generalization
to the non-Abelian case). The quantized version
of the constraint $\nabla{\mathbf E}\approx 0$ reads
\bdm
\frac{\hbar}{\I}\nabla\frac{\delta\Psi[{\mathbf A}]}{\delta {\mathbf A}}
=0,
\edm
where ${\mathbf A}$ is the vector potential.
This equation reflects the invariance of $\Psi$ under spatial gauge
transformations of the form ${\mathbf A}\to{\mathbf A}+\nabla\lambda$. 

The constraints can only be implemented in the form \eqref{WDW} and
\eqref{diffeo} if the quantum version of the constraint algebra
\eqref{algebra-1} -- \eqref{algebra-3} holds without extra c-number
terms on the right-hand side. Otherwise, only a part of the quantum
constraints (or even none) holds in this form. The situation is
reminiscent of string theory where the Virasoro algebra displays such
extra (central or Schwinger) terms. More general quantum constraints
hold there provided the number of spacetime dimensions is restricted
to a specific number (ten in the case of superstrings). It is
imaginable that a restriction in the number of spacetime dimensions
arises also here from a consistent treatment of the quantum constraint
algebra. But so far, this is not clear at all.\footnote{This problem
  was already known to Dirac and was the reason why he abandoned
  working on quantum gravity. In his last contribution to this field,
  he remarked, \cite{Dirac}, p.~543: ``The problem of the quantization
  of the gravitational 
  field is thus left in a rather uncertain state. If one accepts
  Schwinger's plausible methods, the problem is solved. [Dirac refers
  to a heuristic regularization proposed by Schwinger in 1962, C.K.]
  But one cannot be happy with such methods without having a reliable
  procedure for handling quadratic expressions in the
  $\delta$-function.'' Such a reliable procedure is still missing. But
see \cite{feng18}.} 

In the last section, we have seen that we can interpret spacetime as a
generalized trajectory of spaces. In its construction, the four
constraint equations and the six dynamical equations are inextricably
interwoven. What happens in the quantum theory?
There, the trajecory of spaces has disappeared, in the same way as the
ordinary mechanical trajectory of a particle has disappeared in
quantum mechanics. The three-metric $h_{ab}$ and its momentum
$\pi^{cd}$ play the role of the $q^i$ and $p_j$ in mechanics, so it is
clear that in quantum gravity $h_{ab}$ and $\pi^{cd}$ cannot be
``determined simultaneously'', which means that spacetime is absent at
the most fundamental level, and only the configuration space of all
three-metrics respective three-geometries remains. 
This is clearly displayed in Table 1 on p.~248 in \cite{Battelle}.

From this point of view it is clear that in the quantum theory only
the constraints survive. The evolution equations lose their meaning in
the absence of a spacetime. In a certain sense, this is anticipated in
the classical theory by the strong connection between constraints and
evolution equations as discussed in the previous section.

The absence of spacetime, and in particular of time, is usually
understood as the quantum version of the {\em problem of time}. 
It means that the quantum world at the fundamental level is timeless
-- it just {\em is}. Weyl has attributed such a static picture already
to the classical spacetime of GR. 
In \cite{Weyl-Ph} p. 150, he writes:
\begin{quote}
  The objective world just {\em is}, it does not {\em happen}.
  Only from the view of the consciousness crawling upwards in the
  worldline of my life a sector of this world ``lives up'' and passes
  by at him as a spatial picture in temporal
  transformation.\footnote{The German original reads: ``Die 
objektive Welt {\em ist} schlechthin, sie {\em geschieht}
nicht. Nur vor dem Blick des in der Weltlinie meines Lebens
emporkriechenden Bewu\ss tseins ``lebt'' ein Ausschnitt dieser Welt
``auf'' und zieht an ihm vor\"uber als r\"aumliches, in zeitlicher
Wandlung begriffenes Bild.'' [emphasis by Weyl]} 
\end{quote}
In the quantum theory, there is not even a spacetime and a worldline
with a conscious observer, at least not at the most fundamental
level. So how can we relate this picture of timelessness, forced upon
us by a straightforward 
extrapolation of established physical theories, with the standard
concept of time in physics? There are two points to be discussed here.

First, as already mentioned, the DeWitt metric \eqref{DeWitt}
has an indefinite signature: one {\em minus} and five {\em plus}.
This means that the Wheeler--DeWitt equation has a local hyperbolic
structure through which part of the three-metric is distinguished as
an {\em intrinsic timelike variable}. One can show that this role is
played by the ``local scale'' $\sqrt{h}$. In simple cosmological models of
homogeneous and isotropic (Friedmann--Lema\^{\i}tre) universes, this
is directly related to the 
{\em scale factor}, $a$. Using units with $2G/3\pi=1$,  
 the Wheeler--DeWitt equation for a closed Friedmann--Lema\^{\i}tre
 universe with a massive scalar field reads
 \be
 \label{mini}
\frac{1}{2}\left(\frac{\hbar^2}{a^2}\frac{\partial}{\partial a}
\left(a\frac{\partial}{\partial a}\right)-\frac{\hbar^2}{a^3}
\frac{\partial^2}{\partial\phi^2}-a+\frac{\Lambda a^3}{3}+
m^2a^3\phi^2\right)\psi(a,\phi)=0.
\ee
Additional gravitational and matter degrees of freedom\footnote{Except {\em
  phantom fields}, which play a role in connection with discussions
about dark energy, cf. \cite{BKM19,VMS20}.} come with 
kinetic terms that differ in sign from the kinetic term with respect to $a$. For
equations such as \eqref{mini}, one can thus formulate an initial
value problem with respect to intrinsic time $a$. The configuration
space is here two-dimensional and spanned by the two variables $a$ and
$\phi$.

Standard quantum theory employs the mathematical structure of a
Hilbert space in order to implement the probability interpretation for
the quantum state. An important property is the unitary evolution of
this state; it guarantees the conservation of the total probability
with respect to the external time $t$. But what happens when there is
no external time, as we have seen is the case in quantum gravity?
There is no common opinion on this, but it is at least far from clear
whether a Hilbert-space structure is needed at all, and if yes, which
one. This is also known as the {\em Hilbert-space problem} and is
evidently related to the problem of time.\footnote{See e.g.
  \cite{OUP,isrn} for a detailed discussion of this and the other
  conceptual issues discussed below.} 

The second point concerns the recovery of the standard (general
relativistic) notion of time from the fundamentally timeless theory of
gravity. The standard way proceeds via a Born--Oppenheimer type of
approximation scheme, similarly to molecular physics. For this to
work, the quantum state, which is a solution of \eqref{WDW} and
\eqref{diffeo}, must be of a special form. For such a state one
can recover an approximate notion of semiclassical (WKB) time. One can
show that this WKB time (which, in fact, is a ``many-fingered time'')
corresponds to the notion of time in Einstein's theory. Equations
\eqref{WDW} and \eqref{diffeo} then lead to a functional Schr\"odinger
equation describing the limit of quantum field theory in curved
spacetime, the latter given by Einstein's equations. It is in this
limit that one can apply the standard Hilbert-space structure and the
associated probability interpretation.
Higher orders of this approximation allow the derivation of
 quantum-gravitational corrections terms, which, for example, give
 corrections to the Cosmic Microwave Background (CMB) anisotropy
 spectrum proportional to the inverse Planck-mass squared. Such terms
 follow from a straightforward expansion of \eqref{WDW} and \eqref{diffeo} and
 could in principle give a first observational test of quantum
 geometrodynamics, \cite{BKK16}.
  
Quantum geometrodynamics, like practically all approaches to quantum
gravity, is a linear theory in the quantum states and thus obeys the
superposition principle. This means that most states do not correspond
to any classical three-geometry. The situation resembles, of course,
Schr\"odinger's cat. Like there, one can employ the process of {\em
  decoherence} to understand why such weird superpositons are not
observed, see \cite{deco}. Decoherence is the irreversible and
unavoidable interaction of a quantum system with the irrelevant
degrees of freedom of its ``environment''.\footnote{``Environment'' is
 a metaphor here. It stands for other degrees of freedom in
 configuration space which become
 entangled with the quantum system, but which cannot be observed
 themselves.} In quantum cosmology, one can consider, for example, the
variables $a$ and $\phi$ in \eqref{mini} as describing the (relevant)
quantum system, while small density perturbations and tiny
gravitational waves can play the role of the environment. The
entanglement between system and environment leads to the suppression
of interferences between different $a$ and different $\phi$ (within
some limits); in this sense, classical geometry and classical universe
emerge. The same holds for the emergence of structure in the universe
from primordial quantum fluctuations, see \cite{KP09}.

It is evident from the above that the question about the correct
interpretation of quantum theory enters here with its full
power. Since by definition the Universe as a whole is a strictly closed
quantum system, one cannot invoke any classical measurement agent as
acting from the outside. Following \cite{Witt67}, the standard
interpretation used, at least implicitly, is the Everett
interpretation, which states that all components in the linear
superposition are
real.\footnote{Alternatives are the de~Broglie--Bohm approach and
  collapse models, which both are more new theories than new interpretations.}   

It is obvious that at the level of \eqref{mini} there is no
intrinsic difference between big bang and big crunch; both correspond
to the region $a$ approaching zero in configuration space.
This has important consequences for cosmological models in which
classically the universe expands and recollapses, see \cite{Zeh}.
In the quantum version, there is no trajectory describing the
expansion and the recollapse. The only structure available is an
equation of the form \eqref{mini} in which only the scale factor $a$
(and other variables) enter. The natural way to solve such an equation
is to specify initial values on constant-$a$ hypersurfaces in
configuration space and to evolve them from smaller $a$ to larger
$a$. In more complicated models, one can evolve also the entanglement
entropy between degrees of freedom in this way. If the entropy is low
at small $a$ (as is suggested by observations), it will increase all
along from small $a$ to large $a$. There is then a formal reversal of
the arrow of time at the classical turning point, although this cannot
be noticed by any observer, because the classical evolution comes to
an end before the region of the classical turning point is reached.

We have limited the discussion here to quantum geometrodynamics.
The main conclusions also hold for the path-integral approach and to
loop quantum gravity.\footnote{The situation in string theory so far
  is less clear; there are indications that not only the concept of
  spacetime, but also the concept of space is modified, as is
  discussed in the context of the AdS/CFT conjecture.}
In loop quantum gravity, there are analogies with gauge theories, for
example with Faradays's lines of forces, see \cite{FKNR94}.
Still, it is not a gauge theory by itself, and many conceptual issues
such as the semiclassical limit are much less clear than in quantum geometrodynamics.


\section{The role of matter}

Very early on, Einstein was concerned with a fundamental duality
oberved in the physical description of Nature: the duality between
fields and matter. This duality is the prime motivation for
introducing the concept of light quanta
in his important paper on the photoelectric effect
from 1905. At that time, the only known dynamical field was the
electromagnetic field; ten years later, with GR, the gravitational
field joined in.

In {\em Raum, Zeit, Materie}, Weyl writes at the end of the main text,
see \cite{RZM}, p.~317: 
 \begin{quote}
In the darkness, which still wraps up the problem of matter,
perhaps quantum theory is the first dawning light.\footnote{The German
  original reads: ``In dem
  Dunkel, welches das Problem der Materie annoch umh\"ullt, ist
vielleicht die Quantentheorie das erste anbrechende Licht.''}   
\end{quote}
Here, the hope is expressed that quantum theory, which in 1918 was
still in its infancy, may provide a solution
for this duality. This is certainly along the lines of Einstein's 1905
light quan ta hypothesis. But, ten years later, the final quantum
theory gave a totally 
different picture: central notions of the theory are
wave functions and the probability interpretation. Einstein was
repelled by this, especially by the feature of entanglement, which
seems to provide a ``spooky'' action at a distance. This is why he
focused on a unified theory of gravity and electrodynamics. He hoped
to understand ``particles'' as solitonic solutions of field
equations. His project did not succeed. 

A somewhat different direction to understand `matter from space' was
pursued by John Wheeler in the 1950s, see \cite{geon}. The idea is
that mass, charge, and other particle properties originate from a
non-trivial topological structure of space, the most famous example
being Wheeler's wormhole. This is most interesting, but has not led to
anything close to a fundamental theory.\footnote{For a recent account
  of matter from (the topology of) space, see e.g. \cite{Giulini18}.} 

  Weyl's 1929 idea of understanding the interaction of electrons with
  the electromagnetic field by the gauge principle turned out to be more
  promising. The Standard Model of particle physics is an extremely
  successful gauge theory, and virtually all of its extensions make
  use of this principle, too. Gauge fields can also be described in a
  geometric way by adopting the mathematical structure of fibre
  bundles. Still, this is relatively far from the geometric concepts
  of GR, which deal with spacetime and not with the internal degrees
  of freedom of gauge theories. Perhaps gauge theories of gravity may
  help in finding a unified field theory, see \cite{reader}.   
  
Our physical theories all employ a metric to represent matter fields 
and their interactions, so GR is always relevant, even in situations
where its effects are small. As J\"urgen Ehlers writes in 
\cite{Ehlers}, p.~91: ``Since inertial mass is separable from active,
gravity-producing mass, an ultimate understanding of mass can be
expected only from a theory comprising inertia and gravity.''
This should also apply for the origin of the masses in the Standard
Model. The Higgs mechanism provides only a partial answer; the masses
of elementary (non-composite) particles are given by the coupling to
the Higgs, but the masses of composite particles such as proton and
neutron cannot be explained. In fact, it seems that the mass of the proton 
mostly arises from the binding energy of its constituents -- quarks
and gluons -- and not from their masses, which to first order are
negligible. Invoking the inverse of Einstein's famous formula,
$m=E/c^2$, one can speculate that mass ultimately originates from
energy, see \cite{wilczek}. It is hard to imagine that this
origin can be understood without gravity. Perhaps a unified theory at
the fundamental level is conformally invariant, similar to Weyl's 1918
theory, expressing the irrelevance of masses at high energies (small
scales); masses would then only emerge as an effective, low-energy concept.

Unfortunately, despite many attempts, the duality of matter and fields
remains unresolved, even in present approaches to quantum gravity. An
exception may be string theory, but this approach has its own
problems and it is far from clear whether it can be tested
empirically. Perhaps the solution to the problem of matter may arrive
from a completely unexpected direction. Space, time, and matter
continue to be central concepts for research in the 21st century.
The question posed in the title of \cite{Einstein19}, ``Do
gravitational fields play an essential role in the constitution of
material elementary particles?'' will most likely have to be answered
by a definite {\em yes}.


\begin{acknowledgement}

 I am grateful to Silvia De Bianchi and Friedrich Hehl for their
 comments on my manuscript. 
  
\end{acknowledgement}

 \bibliographystyle{harvard}


\end{document}